\def\bbbr{{\rm I\!R}}
\def\lambdabar{\raise 1.7mm \vbox{\hrule width 2mm}\hskip -2mm $\lambda\ $}
\def\ns{nonstandard resonances }
\def\s{standard resonances }

\def\ket#1{{}\mid #1\rangle{}}

\def\obraket#1#2#3{{}\langle#1\mid #2\mid #3\rangle{}}
 \def\dyad#1#2{\mid #1\rangle\langle#2\mid }

\documentclass{ws-ijmpa}

\begin{document}

\vskip 2cm
\title{ RESONANCES FOR A HYDROGENIC SYSTEM OR A HARMONIC OSCILLATOR STRONGLY COUPLED TO A FIELD}

\author{Claude Billionnet}

\address{Centre de Physique Th\'eorique, \'Ecole Polytechnique, CNRS,\\ 91128 Palaiseau, France\\
claude.billionnet@cpht.polytechnique.fr}

\maketitle

\begin{abstract}We calculate resonances which are formed by a particle in a potential which is either Coulombian or quadratic when the particle is strongly coupled to a massless boson, taking only two energy levels into consideration. From these calculations  we derive how the moving away of the particle from its attraction center goes together with the energy lowering of hybrid states that this particle forms with the field. We study the width of these states and we show that stable states may also appear in the coupling.
\end{abstract}

\ccode{PACS numbers: 02.30.Tb, 03.70.+k, 11.10.St, 12.39.Mk}

\section{Introduction}

Our study concerns the resonances which a two-level system forms through its coupling to a massless field. As it is often the case, we introduce simplifications so as to describe the interaction by means of the Friedrichs's model (Ref. 1, Sec. C of Ref. 2 $_{\rm III}$.2), that is to say with a Hilbert space which contains only two kinds of states: on the one hand a state representing the system in its excited state with no boson and on the other hand a continuum representing the system in its fundamental state together with one boson whose state is assumed to depend only on a scalar parameter. The set of resonances which do not come from the energy levels of the decoupled system, resonances which we called elsewhere nonstandard (or of ${\cal C}$-type) depends on the singularities at a finite or infinite distance of the coupling function given by the Hamiltonian matrix elements between the above mentioned two kinds of states. These singularities are not mathematical curiosities; they necessarily exist and in certain cases have a clear physical meaning. The coupling function depends on the wave functions of the two considered levels and its singularities depend on the space extent of these wave functions. If the system is a particle in a potential, these singularities depend on the chosen potential.

It is interesting to study the set of resonances for different potentials. After having studied the square-well case (Ref. 3) we now treat the Coulombian and harmonic potentials when the coupling to the field is strong. Our concern is to store up results on the position of \ns in different contexts in order to understand them better. A particular aspect we focus on is the relation between the width of these resonances and the space extent of the wave functions associated with the two quantum levels of the system. We are also particularly motivated by the following question. Let us consider a particle $q$ bound to its antiparticle $\overline q$, this system interacting strongly with the field of a massless boson. When the field interacts with sufficiently excited $q$$\overline q$ states, that is to say with particles $q$ and $\overline q$ sufficiently apart, can new particles be created, as hybrid states, \ns of zero or very small width. And does this possibility increase with the distance between $q$ and $\overline q$ in one of the levels of the pair? 

The calculation of the resonances in the Coulomb potential case was already tackled in Ref. 4 with the electromagnetic coupling as the coupling to the field. In Sec. 2 we now treat a strong coupling, and larger masses. It makes the \ns pertinent, whereas their widths were extremely large for the electron and its coupling to the transverse photon. We calculate a certain number of them for a certain value of the coupling constant. The harmonic potential case was tackled in Ref. 5 with the same electromagnetic coupling. We are now also going to treat a strong coupling case, showing in Sec. 3 resonances we did not take into account in Ref. 5. Both sections end with a concluding subsection which sums up the main results qualitatively. A synthesis is made in Section 4. Apart from the discussion of the above mentioned question, studies for these two potentials are also a preparation for studying a realistic quark-antiquark potential later, a potential for which computations are more difficult. A first calculation of a nonstandard resonance for such a potential is presented in Ref. 6. Such a future study should take more than two levels into account.

\section{Particle in the  $1/r$ potential strongly coupled to a field}

\subsection{The coupling function and the resonance equation}

Let us reduce the above-mentioned two-particle system to a mass $m$ particle attracted by the origin with the strength 
$\bf F= -\alpha\hbar c\frac{\bf r}{ r^3}$. Besides, this particle is coupled to a transverse massless vector field
$$
{\bf A}({\bf r})=\frac{\hbar^{1/2}}{(2\pi)^{3/2}}\sum_{j=1,2}\int_{-\infty}^\infty \frac{1}{\sqrt {2|{\bf k}|c}}\big({\vec\epsilon}_j({\bf k})\  c_j^*({\bf k})e^{-i  {\bf k}\cdot 
 {\bf r} }+{\vec \epsilon}_j({\bf k})\  c_j({\bf k})e^{i  {\bf k}\cdot 
 {\bf r}}\big)\ d  {\bf k}                                 
$$
through the Hamiltonian 
$$
H_I=-\frac{\sqrt{4\pi\alpha\hbar c}}{m} \ {\bf p}\cdot {\bf A}({\bf r})  \ .\eqno (1)
$$
When the particle is the electron, the force is the attraction that a proton placed at the origin exerts on it and  $H_I$ is the coupling to the transverse electromagnetic field. (The factor $1/\sqrt{\epsilon_0}$, containing the vacuum permittivity, usually present in ${\bf A}$ has been suppressed and transfered into the factor in front of ${\bf p}\cdot {\bf A}$.) In the present study we are going to consider an interaction in which $\alpha$ is no longer  $1/137$ but is of the order of $1$. Besides, the particle is no longer light like the electron but has a mass of the order of the light quarks' one. We are going to show into which resonances the unperturbed levels of the particle in the potential are transformed when the coupling constant changes from $0$ to $1$ (Sec. 2.3.4). But our main concern is to calculate other resonances that the coupling to the field yields.

Let us consider the energy levels of the particle in the $1/r$ potential, levels which are made instable by the interaction  $H_I$. We are interested either in the transition from a state $\ket h$ ( {\it h} for high), with principal quantum number $n_h=2,3,4,\cdots$ and energy  $E_h$, toward a state $\ket l$ ({\it l} for low), with principal quantum number $n_l<n_h$ and energy $E_l$, this transition being accompanied with the emission of a boson of the field, or in the inverse transition and the absorption of the boson. We assume the quantum numbers of the boson are  $j=1,m=0$ and its helicity is $\lambda=+1$. We assume the orbital momentum of state $\ket l$ is $j_l=0$, this entailing taking $j_h=1$. Wanting to reduce the problem to a two-level problem, since this is the only one we know how to treat, we forget all levels except the two we mentioned. Then the states of the particle-field system are in the space spanned by $\ket h$ and $\ket l\otimes\varphi(k)$, where $\varphi$ is in $L^2(\bbbr,dk/k)$. The Hamiltonian acting in this space is
$$
H=\big(E_l\dyad ll+E_h\dyad hh\big)\otimes 1+1\otimes H_{\rm rad}+\dyad hl\otimes c(\overline g_I)+\dyad lh c^*(g_I)                            \eqno (2)
$$
with
$$
g_I(k):=\obraket h{H_I}{l,k}=i\pi^{-1/2} a\ \hbar c\ \alpha^{3/2}\ \Phi (k) \eqno (3)
$$
where $a=\hbar/(mc\alpha)$ and
$$
\Phi(k)=\int_0^\infty j_1(kr) R_h^*(r)R_l'(r)r\ dr\ .
$$
(See Ref. 7.) The $R(r)$'s are functions of $r/a$. 

\noindent
We set $R^{\rm red}(\rho):=a^{3/2}R(\rho a)$, $\Phi^{\rm red}(y):=a^2\Phi(y/a)$ and $g_I^{\rm red}(y):=E_{hl}^{-1}g_I(y/a)$; then
$$
g_I^{\rm red}(y)=i\pi^{-1/2}\  \frac{\hbar c}{a}\alpha^{3/2}\ E_{hl}^{-1}\Phi^{\rm red}(y)= i\pi^{-1/2}\alpha^{5/2}\ \frac{mc^2}{E_{hl}}\ \Phi^{\rm red}(y)\ .               \eqno (4)
$$

\noindent
Set $E_{hl}=E_h-E_l$. The energy $z$ of the resonances is given by  $z=E_l+E_{hl}\ \zeta$, where $\zeta$ is a zero of the analytic continuation into the second sheet of the function
$$
f(\zeta):=\zeta-1-2\int_0^\infty\frac{|g_I^{\rm red}(y)|^2}{ \zeta-\alpha\frac{mc^2}{E_{hl}}y}\ \frac{dy}{y}\ .
$$  
Setting $\mu=\alpha mc^2/E_{hl}$, and using $E_{hl}=({1}/{n_l^2}-{1}/{n_h^2})\ \alpha^2\ mc^2/2$, we can write $f$ as
$$
f(\zeta)=\zeta-1-\frac{8}{\pi}\ (\frac{1}{n_l^2}-\frac{1}{n_h^2})^{-2}\ \alpha\int_0^\infty\frac{|\Phi^{\rm red}(y)|^2}{\zeta-\mu y}\frac{dy}{y}       \eqno (5)
$$
The analytic continuation $f_+$ of this function into the lower half-plane, across $\bbbr^+$, is given by
$$
f_+(\zeta)=\zeta-1-\frac{8}{\pi}\ (\frac{1}{n_l^2}-\frac{1}{n_h^2})^{-2}\ \alpha\int_0^\infty\frac{|\Phi^{\rm red}(y)|^2}{\zeta-\mu y}\frac{dy}{y}+16\  i\ (\frac{1}{n_l^2}-\frac{1}{n_h^2})^{-2} \zeta^{-1}\alpha\ \Phi^{\rm red}(\frac{\zeta}{\mu})^2 \eqno (6)
$$
The key point in our study, a point our preceding works were focused on, is that the singularities of $\Phi^{\rm red}$ are the cause of the existence of resonances which do not tend to $E_h$, the energy of the excited state, when the coupling tends to $0$. They are the nonstandard resonances. We get here the precise meaning of the expression "do not come from" used as a shortening in the first paragraph of the Introduction. The pole of $\Phi^{\rm red}$ entails a pole in $\zeta$ of the integral and hence of $f_+$, and, when $\alpha$ is small, the zeros of  $f_+$, corresponding to the nonstandard resonances, are close to this pole of the integral.

\subsection{The singularities of the coupling function and their interpretation}

 For the system we consider here, we showed (Ref. 4) that the singularities of $\Phi^{\rm red}$ in the lower half-plane reduce to one pole. (It must be noted, however, that its multiplicity varies with the pair of levels which is considered.)

The imaginary part of this pole is related to the space extension of states $\ket h$ and $\ket l$. More precisely, the two wave functions of $\ket h$ and $\ket l$ decreasing at infinity like $e^{-r/r_i}$, $i=h,l$, let $\overline\rho$ denote half the harmonic mean of distances $r_h$ and $r_l$, in units of \lambdabar$\!_C$, $1/2\pi$ times the particle's Compton wave-length. The imaginary part of the pole is then  $(\overline\rho)^{-1}$ (See Ref. 4.) We have $\overline\rho=n_hn_l/(n_h+n_l)$. The more extended the pair, hence the more excited its states, the closer to the real axis the


\begin{table}[ph]
\tbl{Correlation between the pole of the coupling function and the extent of the pair of levels}
{\begin{tabular}{@{}ccccccc@{}} \toprule
 pair & (2,1) &(3,1)  &(4,1)  &(3,2)  &(4,2)  &(4,3) \\ \colrule
$\overline\rho$& 0.67 &0.75&0.80&1.20&1.33&1.71 \\ 
 pole of $\Phi^{\rm red}$ &-1.50i&-1.33i&-1.25i&-0.83i&-0.75i&-0.58i \\ \botrule
\end{tabular} \label{ta1}}
\end{table} 


\noindent
pole of $\Phi^{\rm red}$ (see Table 1), and therefore the closer to the real axis the resonances if $\alpha$ is small. (The \ns nevertheless remain far from the reals for the atom in the electromagnetic field; see below.) If $\alpha$ is not small, for instance if it is of the order of $1$, a precise study must be carried; this is the aim of this Sec. 2.

Denoted by $C$ in $mc^2$ units, the energies corresponding to these poles in $\zeta$ are
\vskip -0.2cm
$$
C=-\frac{1}{2} n_l^{-1}\alpha^2+\alpha\  ({\rm pole\  of\  \Phi^{\rm red}})\ .\eqno (7)
$$
\vskip -0.2cm
For pair $(2,1)$ and $\alpha=1/137$, we approximately recover the width of the \ns we calculated in Ref. 4 in the electron case. Indeed, we have: $|\Im(C)|=1.5/137\times 511$ keV $\simeq 5.6$ keV, the width of the \ns being $5.55,\ 5.57$, $\ 5.60$ and $5.62\ $ keV. The width of level $2$ is approximately $2\times10^{-7}$ eV (if one considers one helicity only), which makes a factor of $3\times10^{10}$ between the two widths.

In contrast, for a mass of $220$ MeV, $\alpha=1$, and pair (2,1), the absolute value of the imaginary part of the pole of the resolvent's matrix element is $1.5\times 220$ MeV$=330$ MeV, a value much more interesting than that of the electromagnetic case.

\subsection{Computation of the resonances for $m=220$ MeV and $\alpha=1$, with several level pairs}

For information only, let us give some coupling functions (the referred pair is indicated in the subscript):
$$\displaylines{
\Phi^{\rm red}_{21}=16\sqrt\frac{2}{3}\ y\frac{1}{(9+4 y^2)^2}\ ,\quad\Phi^{\rm red}_{31}=16\sqrt{6}\ y\frac{8+9y^2}{(16+9 y^2)^3}\cr\Phi^{\rm red}_{32}=-2^7\sqrt{3}\ y\ \frac{125+216y^2(-1+6y^2)}{25+36y^2)^4}
}$$
Fig. 1 gives the result of the numerical computation of the \ns for the first six pairs of levels. They are shown by bold points in the complex energy plane. The energy unit on the axes is $mc^2$. The level energies in this unit are given in Table 2. The real part of $C$ gives the energy of the lowest level of the pair. The absolute value of the resonances' imaginary parts  are to be interpreted as usual as their widths.

\vskip 0.7 cm


\begin{figure}[h]
\centerline{\psfig{file=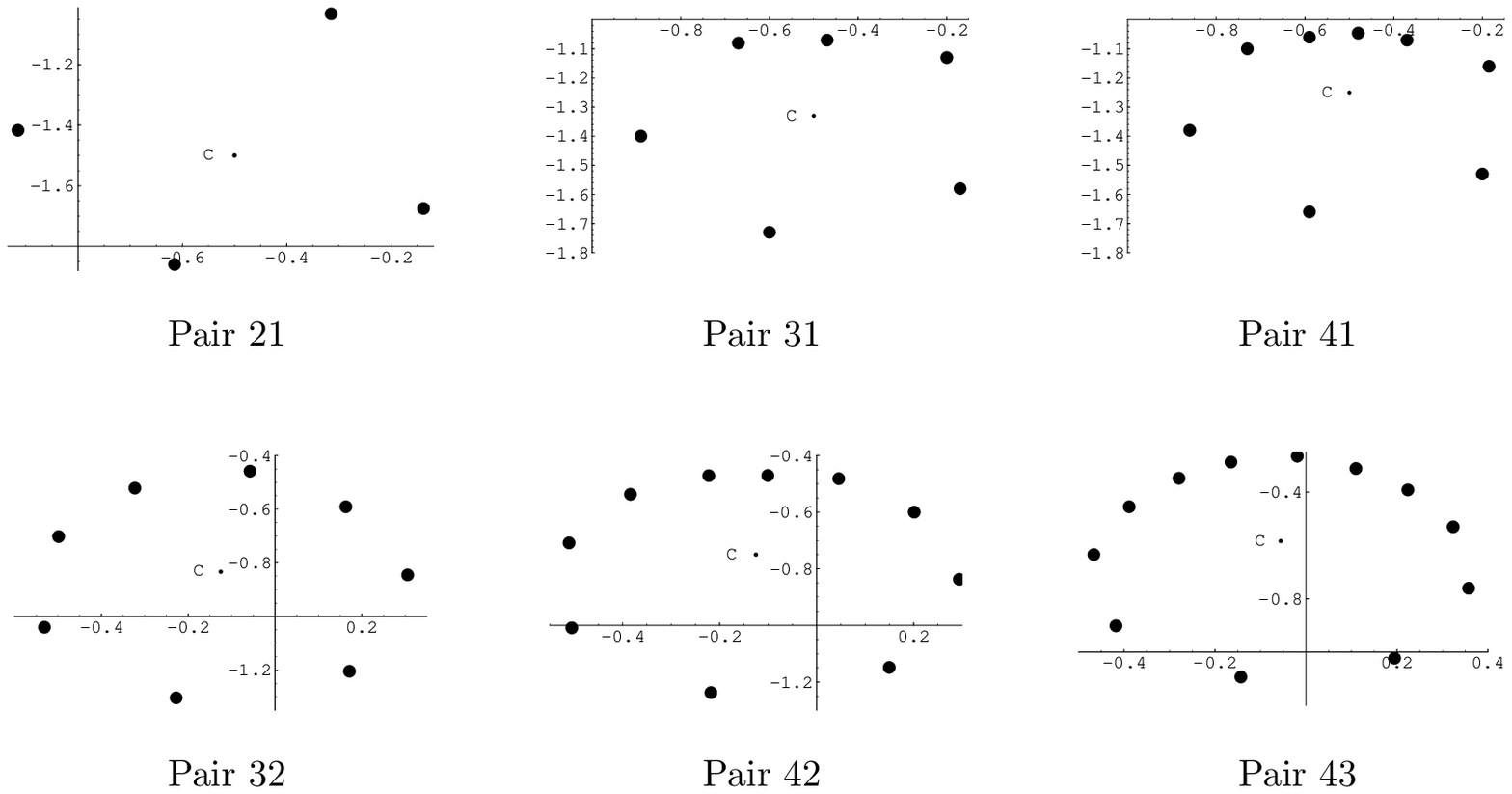,width=12cm}}
\vspace*{8pt}
\caption{Variations of \ns with the excitation of the pair.\label{f1}} 
\end{figure}


The number of resonances increases with the excitation of the pair (because the order of the pole of $\Phi^{\rm red}$, and therefore the order of that of $f_+$,
 increases) and their energies are disposed around the value $C$, represented by a finer point in the figure. Resonances being all the more interesting when their width is small, we are going firstly to focus on these widths and particularly on their dependence on the space extent of the states, through $\Im(C)$. 

\subsubsection{Width of the \ns}

When the extent of the pair increases, the fact that the pole of $\Phi^{\rm red}$ comes closer to the reals manifests itself in $C$ coming closer to the reals. Some resonances have a width greater, others a width smaller than $\Im(C)$. We leave the problem of the physical meaning of each resonance aside and we focus on the resonance with the smallest width. Its value together with $\overline\rho$, the mean extent of the pair, is given in Table 2. $\kappa$ represents the strength of the coupling, in $E_h-E_l$ units, i.e. $\kappa=\big(2\int_0^\infty |g_I^{\rm red}(y)|^2dy/y\big)^{1/2}$.


\begin{table}[ph]
\tbl{For each pair, energy of the two levels, energy and width of the narrowest nonstandard resonance, space extent and relative coupling strength}
{\begin{tabular}{@{}ccccccc@{}} \toprule
 pair & (2,1)&(3,1)&(4,1)&(3,2)&(4,2)&(4,3) \\ \colrule
$E_h/mc^2$&-0.125&-0.056&-0.031&-0.056&-0.031&-0.031 \\
$E_l/mc^2$&-0.5&-0.5&-0.5&-0.125&-0.125&-0.056 \\
$(E_h-E_l)/mc^2$& 0.375& 0.444& 0.469& 0.069& 0.094& 0.025 \\
energy/$mc^2$&-0.32&-0.47&-0.48&-0.06&-0.10&-0.02 \\
width/$mc^2$& 1.03& 1.07 & 1.05& 0.46& 0.47& 0.26 \\
$\overline\rho$ & 0.67 & 0.75& 0.80& 1.20& 1.33& 1.71 \\
$\kappa$& 0.079& 0.042& 0.027& 0.020& 0.013& $7\times 10^{-3}$ \\ \botrule
\end{tabular} \label{ta2}}
\end{table}


\noindent
The width for pairs (2,1), (3,1), (4,1) is about $330$ MeV; it falls down to $102$ MeV for pairs (3,2) and (4,2) which have a greater space extent and to $57$ MeV for pair (4,3) the space extent of which is still greater. But one has to take into account the fact that the level spacing decreases so that the ratio of the widths to this level spacing in fact increases: widths are greater than  $E_h-E_l$ by a factor of 2 or 3 for pairs based on level 1, a factor of 5 or 7 for those based on level 2 and a factor of 10 for level 3. This limits the physical meaning that can be given to these states, for the considered $\alpha$ value.

Let us consider pairs $(n,1)$ and examine whether narrow resonances could be obtained through making $n$ large. The extent $\overline\rho$ is $1-1/(n+1)$ and its upper limit is $1$ if $n$ goes to infinity. The pole of $\Phi^{\rm red}$ approaches the reals up to the limit value  $\zeta=-i$, corresponding to $|\Im(C)|=1$. The ratio $|\Im(C)|/E_I$ is $2$. Therefore, the answer is no. For the series $(2,n)$, the limit of $\overline\rho$ is $2$ and the pole of $\Phi^{\rm red}$ approaches the reals up to $-i/2$, which corresponds to $\Im(C)=-{1}/{2}$. The width does decrease when we change from $(n,1)$ to $(n,2)$, but the ratio to the level spacing increases: it is now $4$. We see that we never get small ratios. Therefore, as regards the question mentioned in the Introduction about the creation of narrow states for large $n$, we see that the $1/r$ potential does not provide an interesting setting.

When both principal quantum numbers increase, $C$ approaches the reals. The calculation gets more and more difficult, if only because of the increasing number of resonances. However we can get an idea of the displacement of the cluster in the case where $n_h$ becomes large and  $n_l=n_h-1$. We saw that $\Im(C)=-(n_h+n_l)/(n_hn_l)\  \alpha\ mc^2$. For $n_l=n_h-1$ and large $n_h$, we have
$
|\Im(C)|\simeq 2n_h^{-1}\alpha\  mc^2.
$
For large $n_h$, $|\Im(C)|$ may become quite small. For instance, with  $m=220$ MeV, and $n_h=50$, $|\Im(C)|\simeq 8.8$ MeV and for $n_h =300$, $|\Im(C)|\simeq 1.5$ MeV. Therefore, we can expect narrow resonances. But the difference between the energies of the two considered levels also becomes small, decreasing as $n_h^{-3}$. As a consequence the ratio of the width to the level spacing increases like  $n_h^2$. Hence the resonances will not be seen for large $n_h$, in our two-level approximation.

Nevertheless, for larger $\alpha$ values, the resonances' width may be small, as Fig. 2 shows it. For pair (2,1), the figure shows the energy and width (vertical line) of the narrowest nonstandard resonance for different $\alpha$ values. (The unperturbed levels are drawn in bold lines.) Since the "ionization energy" increases like  $\alpha^2$ (the potential is proportional to $\alpha$) and $|\Im(C)|$ behaves like $\alpha$, we already expect the ratio of the widths to the level spacing to decrease when $\alpha$ increases. But the diagram shows that the width itself decreases. Besides, the resonance's energy approaches the lowest level of the pair. The resonance even becomes real negative, that is to say an eigenvalue of the Hamiltonian, if $\alpha\geq 5.54$.


\begin{figure}[h]
\centerline{\psfig{file=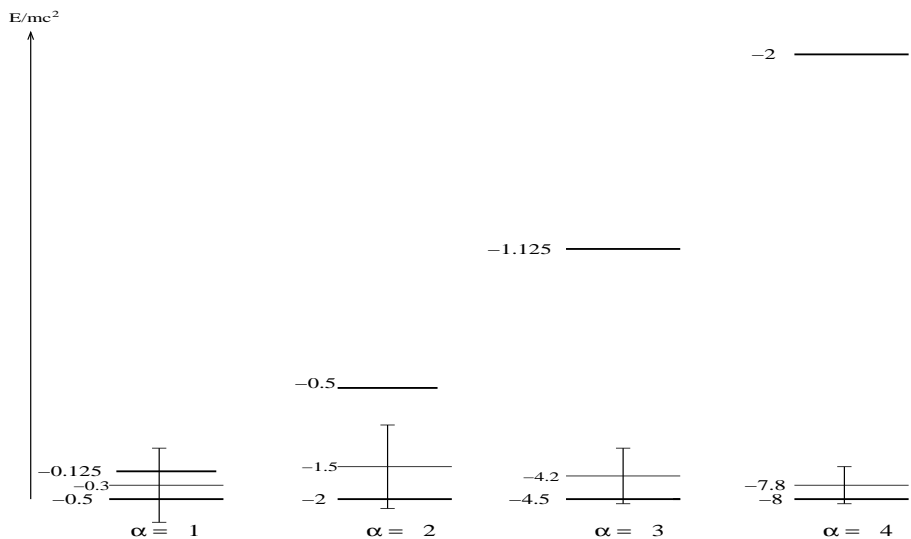,width=12cm,height=7cm}}
\vspace*{8pt}
\caption{Variation of a nonstandard resonance with the coupling, for pair (2,1).\label{f2}} 
\end{figure}


In order to see what role \ns actually play in strong interactions, it is thus important to know what coupling constant should be used.

It is also necessary of course to know what potential is the most appropriate. In this respect, let us mention the result of a calculation for the two lowest levels in a model where two $220$ MeV masses attract each over through a potential of the form  $a/r+b r$, this system being coupled to the field of a massless vector boson, with a coupling constant $\alpha=0.6$: we find a nonstandard resonance of $(485-390\ i)$ MeV, whose width is only $0.6$ times the energy difference between the two uncoupled states at $E_h=1370$ MeV and $E_l=730$ MeV. (There is an error in Ref. 3 due to an extra factor of 2 in the coupling $\lambda_0$.) Hence there are physically reasonable potential for which the \ns are more interesting than those obtained with the $1/r$ potential and $\alpha=1$. We are going to see what happens in the harmonic oscillator case in Sec. 3.

For pair $(2,1)$, we noted that, if the coupling is strong enough, the width decreases when the coupling increases, becoming even $0$. $\Im(C)$ also decreases at fixed $\alpha$, for pairs $(n,1)$, if $n$ increases. Hence one could look for narrow widths for large $n$ and large $\alpha$, the latter not too large to keep a physical meaning. For example: do we find narrow resonances for pairs $(n,1)$ with a coupling  $\alpha=2$ or $3$? The computation shows that the answer is no: for pair $(3,1)$, the width increases with $\alpha$ for  $\alpha=1,2,4$ (whereas it decreased for pair $(2,1)$ for $\alpha=2,3,4$). The width starts decreasing only for greater values of $\alpha$. It vanishes for $\alpha>11.89$. We recall that the critical value for pair $(2,1)$ was $5.54$.

The space extent of the levels does not provide easily measurable \ns although their width are much smaller with respect to the level spacing than for the electron and the electromagnetic coupling.

Up to now we have been interested in the widths of the nonstandard resonances, widths which are closely related to the poles of $\Phi^{\rm red}$. Let us now focus on how the resonances' energy vary when the particle moves away from the origin.

\subsubsection{Lowering of the energy of certain \ns when the particle moves away from the origin}

For pairs (2,1), (3,1) and (4,1) on the one hand and for pairs (3,2) and (4,2) on the other hand, Table 2 shows a lowering of the energy of the narrowest resonance down toward the lowest energy of the pair, whereas the widths are almost constant in the two series.  Thus making the distance of the particle from the attraction center (or the distance between two particles) larger when considering successively states 2,3 and 4 or states 3 and 4 lowers the energy of one of the \ns associated with the transition from the fundamental state to these states.
This property is noticeable, although it would be more interesting if the resonances were narrower. But we will see this property is also true for the system studied in Sec. 3. In the following section we show that the property holds for $\alpha=0.5$, whereas the situation is more complicated for $\alpha=2$.

\subsubsection{Results for  $\alpha=2$ and  $\alpha=0.5$}

For $\alpha=2$, which gives $E_2=-2$, and still for the narrowest resonance, the gap to the fundamental does not decrease but increases when we shift from pair $(2,1)$ to pair $(3,1)$: it goes from $0.45$ to $0.7$ (the width also increases, going from  $1.1$ to $1.88$, whereas the two widths were almost equal for $\alpha=1$). But for pair $(3,1)$, there is a resonance with a lower energy and a width only slightly greater than the one we just considered: this resonance is circled in Fig. 3 which gives all the \ns  for pair (3,1), in the complex energy plane and in $mc^2$ units.
This resonance is $0.04$ above $E_2$, this time closer to $E_2$ than the narrowest resonance for pair (2,1). We would then recover the lowering. However we see that the great number of \ns  complicates the qualitative analysis of the influence of the particle's position on the energy of the resonances. For pair (4,1), and still for $\alpha=2$, there is a resonance with about the same energy as the one we just mentioned (and also the same width $\simeq 2.02$).


\begin{figure}[h]
\centerline{\psfig{file=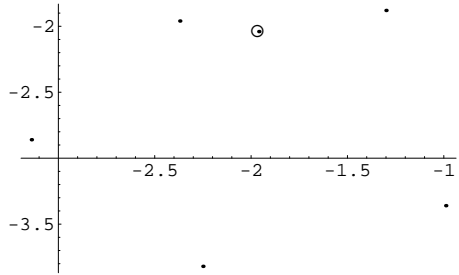,width=6cm}}
\vspace*{8pt}
\caption{Nonstandard resonances  for pair (3,1) with $\alpha=2$.\label{f3}} 
\end{figure}


On the contrary, the phenomenon we saw for $\alpha=1$ repeats itself for $\alpha=0.5$, since the gap between the narrowest resonance and the fundamental decreases, going from  $0.053$ to $0.025$ when we change from pair (2,1) to pair (3,1). (The width also decreases, going from  $0.62$ to $0.56$, the level spacing being $0.19$ for (2,1).)

In any case, for the three values of $\alpha$ we considered, the interest of the resonances seems limited by their large widths.

\subsubsection{Comparison between the widths of standard and \ns}

The perturbed levels' energies are close to
the unperturbed ones' and we do not give them here. As regards the width $\Im(z_s)$ ({\it s} for standard) that the upper level of the pair acquire through the coupling to the field, its ratio to $E_h-E_l$ is given in Table 3.

\vfill\eject


\begin{table}[ph]
\tbl{Width of the nonstandard resonances for different pairs}
{\begin{tabular}{@{}ccccccc@{}} \toprule
pair & (2,1)) & (3,1) & (4,1) & (3,2) & (4,2) & (4,3)  \\ \colrule
$\Im(\zeta_s)$ & 0.04 & 0.009 & 0.004 & 0.009 & 0.003 & 0.004 \\ \botrule
\end{tabular} \label{ta3}}
\end{table} 


\subsection{Conclusion}

For pair (2,1), $\alpha=1$ and $m=220$ MeV, the energies of the two levels being $E_l=-110$ MeV and $E_h=-27.5$ MeV, the energy of the narrowest nonstandard resonance is $E=(-70.4-226.6\ i)$ MeV. The width is large but not very large.

Increasing the coupling sufficiently allows to reduce the width of nonstandard resonances, but we need quite strong couplings to change these resonances into stable states. For example  $\alpha=5.54$ for pair (2,1), and $\alpha=11.89$ for pair (3,1).

Going from level 2 to levels 3 and 4, therefore increasing the distance between the particle and the attraction centerer, lowers the energy of the narrowest nonstandard resonance toward level 1, for $\alpha=1$ or $\alpha=0.5$. However, its width is not small.

Considering pairs of very excited levels yield \ns whose widths are small, for example $8$ MeV, but large compared to the level spacing, a spacing that decreases as the excitation increases.

We are now going to consider a system for which the excitation of the levels does not go together with their narrowing. In this case, the influence of the space extent of the uncoupled states on the energy and width of the \ns will be more perceptible.

\section{Harmonic oscillator strongly coupled to a field}

\subsection{Coupling function and resonance equation}

Let us consider a one-dimensional harmonic oscillator with mass $m$ and frequency $\omega$. Les us assume it is coupled to the field of a massless scalar boson whose states are described in $L^2(\bbbr)$. We assume the only non-zero matrix elements of the interaction Hamiltonian $H_I$ are
$$
\obraket{\psi_m\otimes\varphi}{H_I}{\psi_n\otimes\Omega}:=\lambda\int g_{nm}(k)\varphi(k) dk                               \eqno (8)
$$
where $g_{nm}(k)=\int\overline\psi_m(x)\psi_n(x) e^{-ikx}dx$. $\psi_n$ denotes the wave function of the $n$$^{\rm th}$ level above the fundamental one and $\Omega$ denotes the field vacuum. $\varphi$ is any boson state,  $k$ the wave number. The factor $\lambda$ allows us to make the strength of the coupling to the field vary; its dimension is that of an energy times the square root of a length.

Note that we do not set ourselves here in the frame of the Wigner-Weisskopf model (see for instance  Ref. 8, or Ref. 9). The latter excludes transitions between states the quantum numbers of which differ by more than one and it only involves one coupling function $g$. The coupling functions we choose here have some analogy with those derived from the coupling of the atom to the electromagnetic field. In this latter case, one of the wave function occurs with a derivative. The particular $H_I$ we take here is supposed to provide information useful for other interaction models.

Let us select two states of the oscillator: $\ket n$ and $\ket m$, with $n>m$, and consider the Hamiltonian
$$\displaylines{
H_{nm}:=(E_n\!\dyad nn+E_m\!\dyad mm)\otimes 1+1\otimes H_{\rm bos}+ \dyad nm\otimes c(g_{nm})+\hfill\cr\hfill\dyad mn\otimes \big(c(g_{nm})\big)^*
}$$
$H_{\rm bos}$ is the energy operator in the bosonic space. The eigenvalue equation is
$$
z-E_n-\lambda^2\int\frac{|g_{nm}(k)|^2}{z-E_m-\hbar c |k|}dk=0    \eqno (9)
$$
Let us go to dimensionless quantities through setting $\zeta=(E_n-E_m)^{-1}(z-E_m)$ and $y=\delta k$, with $\delta=\sqrt{\hbar/m\omega}$. (The wave functions decrease at infinity like  $e^{-\frac{1}{2}(\frac{x}{\delta})^2}$.) Let $\mu=c/(\omega\delta)$ denote the ratio between $(2\pi)^{-1}$ times the wavelength of the transition between the first two levels and the extent of the wave functions measured by $\delta$. Set $G_{mn}(y):=g_{mn}(\delta^{-1}y)$. In variable  $\zeta$, (9) becomes, with $E_{nm}:=E_n-E_m$ and $\mu_{nm}=\hbar c/ (E_{nm}\delta)=(n-m)^{-1}\mu$:
$$
f(\zeta)=\zeta-1-\frac{\lambda^2}{\delta E_{nm}^2}\int\frac{|G_{nm}(y)|^2}{\zeta-\mu_{nm}|y|}dy=0                            \eqno (9')
$$
Resonances are given by the zeros of the analytic continuation of $f$ into the lower half-plane, across  $\bbbr^+$, a continuation given by
$$\displaylines{
f_+(\zeta)=\zeta-1-\frac{2\alpha^2}{(n-m)^2}\int_0^\infty\frac{|G_{nm}(y)|^2}{\zeta-\mu(n-m)^{-1}y}dy+\hfill\cr\hfill(-1)^{n+m}\frac{4 i \pi \mu^{-1}\alpha^2}{n-m}G_{nm}\big(\mu^{-1}(n-m)\zeta\big)^2\quad (10)
}$$ 
where $\alpha=\lambda/(\sqrt{\delta}\hbar\omega)$. We used that $|G(y)|^2$ is even.

In the domain of high energy physics, we choose $\delta=1$ Fermi and, so as to get $\mu=1/2$, we take $m=98,6$ MeV. We then have $\hbar\omega=395$ MeV.

\subsection{Results concerning \ns}

The coupling functions $G_{nm}$ have the following form 
$$
G_{nm}(y)=y^rP_{nm}(y^2)\ e^{-y^2/4}\eqno (11)
$$
where $r$ is  $1$ or $2$ according to whether $n+m$ is even or odd. Hence they tend to infinity when $y$ goes to infinity in certain sectors of the complex plane. As a consequence, there exist zeros of $f_+$ which do not tend to $1$ when $\alpha$ goes to $0$, these zeros thus corresponding to nonstandard resonances.

Let us give the results of the computation of some of these resonances, limiting sometimes ourselves to a qualitative description. We have limited ourselves to  $n\leq 8$, without computing the resonances for all pairs.

\subsubsection{A first type of nonstandard resonance: $z_0$}

First and foremost, there are nonstandard {\it eigenvalues}, that is to say eigenvalues which do not tend to the energy $E_n$ of the upper state when  $\alpha$ tends to $0$. This is the case for all pairs $(n,n-1)$. To each eigenvalue corresponds a mixed oscillator-field state which is an eigenvector of Hamiltonian $H_{nm}$. Its energy is below $E_m$. For instance, for pair $(1,0)$, the eigenvalue is $85$ MeV, below $E_0=197.5$ MeV, and for pair $(6,5)$, it is $2106$ MeV, below $E_5=2172.5$ MeV. If $\alpha$ decreases below a certain value (depending on the pair), each of these eigenvalues becomes a resonance with a certain width. Generically, $z_0$ will denote a resonance which becomes real if $\alpha$ increases. We begin with describing resonances of this latter type for pairs not of the form $(n,n-1)$, except for $n=1$, and still for $\alpha=1$.

For pairs of the form $(n,0)$, the width of these resonances $z_0$ increases with $n$ but stays small for the first $n$'s: successively $0$ for pair $(1,0)$ (see above), then $\ 13\  {\rm MeV}$ for $n=2$, $\ 55\  {\rm MeV}$ for $n=3$, $\ 95 \  {\rm MeV},\ 124\  {\rm MeV}$ etc.. These resonances $z_0$ for


\begin{figure}[h]
\centerline{\psfig{file=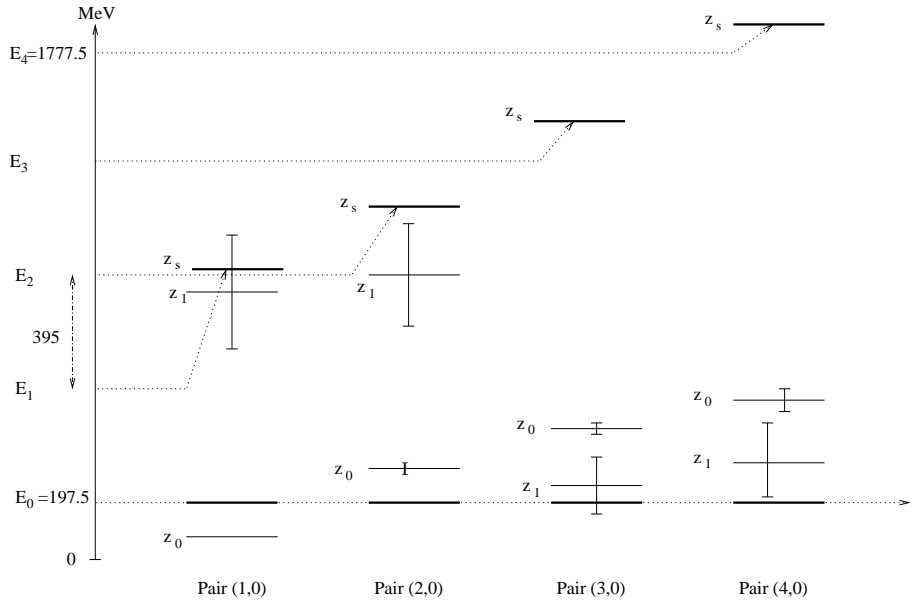,width=12cm}}
\vspace*{8pt}
\caption{Some resonances for the oscillator strongly coupled to a field.\label{f4}} 
\end{figure}


\noindent
pairs $(n,0),n=1,2,3,4$,  are represented in Fig. 4, which gives the lowest energy resonances together with their widths, for these pairs. The lowest level $E_0$ is represented in bold lines. The standard resonances also; their widths are smaller than  $4$ MeV. When they are not zero, the widths of the other resonances are represented by a vertical line.
For the series $(n,1)$, the width of resonances $z_0$ also increases with $n$ and stays small: widths are respectively $0$ for pair $(2,1)$, then $13\  {\rm MeV},\ 50\  {\rm MeV},\ 81 \  {\rm MeV},\ 105\  {\rm MeV}\ \cdots$.

In order to relate the energy $\Re(z_0)$ of these resonances to the extent of the level pairs, it is convenient to have a number measuring this extent at our disposal. Indeed, contrarily to what happened for the $1/r$ potential of Sec. 2, with the mean extension $\overline\rho$ (derived from the $e^{-r/r_i}$ behavior of the wave functions), here the parameter $\delta$ which measures the wave-function extent does not vary with the levels. So as to measure more precisely the space extent of each pair, let us introduce the number $d_{(n,m)}:=2(d_n^{-1}+d_m^{-1})^{-1}$ where $d_n=(\int_0^\infty x^2 \psi_n(x)^2 )^{1/2}$. The results of the computation, some of which being illustrated with Fig. 4, show that $\Re(z_0)$ increases with $n$, as does the gap between the two levels. But the ratio $(\Re(z_0)-E_0)/(E_n-E_0)$ between these two numbers decreases for 
$n=5,6,7,8$. It decreases only slowly, staying in the neighborhood of  $0.22$; this may be put in relation with the fact that $d_{n,0}$ increases only slowly when $n$ increases  ($d_{2,0}=0.98,\ d_{80,0}=1.31$). The same is true for the family $(n,1)$. We are now going to consider resonances for which the gap to the lower level, not only its relative value, decreases as the extent increases.

\subsubsection{A nonstandard resonance of a second type: $z_1$. Lowering of the energy when the distance of the particle to the center increases}

$z_0$ mentioned above is not the only nonstandard resonance. For each pair, we have an infinite number of other resonances. Let us denote them by $z_1,\ z_2,\ \cdots$, arranged in increasing order of their real parts. As an example, Fig. 5 shows the variation with $\alpha$ of the lowest energy resonances for pairs $(2,1)$ and $(6,1)$. They are expressed in the $\zeta$ variable. One sees how these resonances tend to infinity when $\alpha$ tends to $0$. For pair (2,1) the branch $\zeta_0$ becomes real for  $\alpha\geq 1/\sqrt 2$, yielding the above mentioned eigenvalue for $\alpha=1$.
 

\begin{figure}[h]
\centerline{\psfig{file=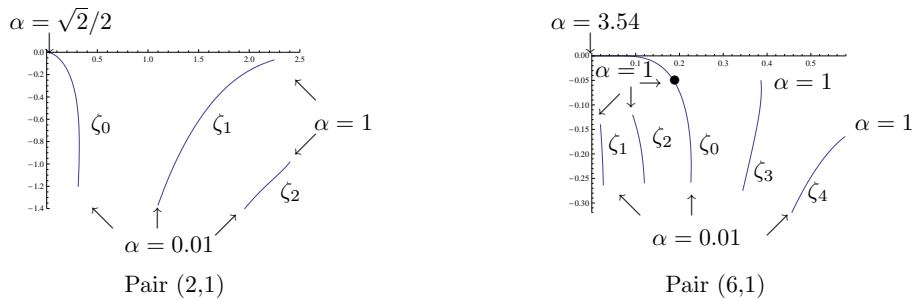,width=12cm}}
\vspace*{8pt}
\caption{Variation of different \ns when $\alpha$ varies, in the $\zeta$-plane.\label{f5}} 
\end{figure}


Let us focus on $z_1$, the resonance with the lowest energy (excluding possibly $z_0$). For the series $(2n+1,0)$, its energy decreases when $n$ successively takes the values $0,1,2,3$, approaching the energy $E_0$ of the lowest state (see Fig. 4 for $n=0,1$), whereas the width, quite large ($>200$ MeV), increases for $n>0$. The energy also decreases for the series $(2n,0)$, $n$ varying from $1$ to $4$ (see Fig. 4 for $n=1,2$), 
whereas the width increases for  $n>1$. We find the same characteristics for the pairs $(n,1)$, $n$ even or odd. The large width of the resonance may be an obstacle to a physical interpretation of this lowering of the energies. However, as can be seen in Fig. 4, this width is only a fraction of the gap between the two levels. For pair $(7,0)$, the width of the  resonance with the lowest energy above $E_0$ is $401$ MeV, but it is seven times less than the energy difference between the two levels ($2765$ MeV). Later on, we are going to see that the same lowering phenomenon occurs for pairs $(n,n-1)$ but this time with narrow widths. Let us note here incidentally that there is a resonance with a small width of $23$ MeV for pair $(3,1)$. (We give these figures just to point out the possible importance of some nonstandard resonances in a concrete way; we of course do not claim here, nor elsewhere in this paper, to any precise realistic description of strong interactions.)  

It has to be noted that resonances with energies greater than $z_1$'s may also have small widths: for instance, for pair  $(4,1)$, $\Im(z_2)=52$ MeV.

For the family $(n,n-1)$, let us consider the lowest energy resonance once we have put the already mentioned eigenvalue aside. Its energy lowers toward $E_{n-1}$ when $n$ increases. Indeed, for pair (2,1), the energy is $1477$ MeV, $885$ MeV above $E_1$,  and the width is $27$ MeV; for $(3,2)$, the energy is  $1588$ MeV, that is to say $600$ MeV above $E_2$, and the width is $0.4$ MeV; for $(4,3)$, the energy is $1679$ MeV, $296$ MeV above $E_3$, and the width is $10$ MeV. We see the widths are small, which gives the energy of these resonances more meaning. The result can be expressed through saying that the gap to the lower level is a decreasing function of the above defined pair's mean extent, for $n\geq 2$. Table 4 gives this gap in MeV as a function of $d_{(n,n-1)}$ (in Fermi), for various $n$'s.


\begin{table}[ph]
\tbl{The gap to the lower level and the width of one of the nonstandard resonances as a function of the space extent}
{\begin{tabular}{@{}cccccccc@{}} \toprule
(n,n-1) & (1,0)&(2,1)&(3,2)&(4,3)&(5,4)&(10,9)&(20,19) \\ \colrule
$d_{(n,n-1)}$ & 0.90 & 1.38& 1.71& 1.99& 2.23& 3.16& 4.49 \\
$E-E_{n-1}$& 735 & 885& 600& 296& 266 & 189 & 134  \\ 
width & 395 & 27& 0.4& 10& 11 & 10& 9 \\ \botrule
\end{tabular} \label{ta4}}
\end{table} 


\subsection{The \s}

As the standard resonance is concerned, widths are of the order of a few MeV for pairs $(n,n-1)$, $n$ varying from $1$ to $6$, or for pair $(3,1)$. We thus note the widths of the \s are comparable, for these pairs, to those of the above mentioned nonstandard resonances. The widths of the \s become smaller and sometimes very small for pairs $(3,0),\ (4,0),\ (5,0),\ (6,0)$, or $\ (4,1),\ (5,1),\ (6,1)$. Contrary to what might be expected, the reason why the widths are so small is not that the strength of the coupling gets small: for example we have $||g_{60}||=0.75$, whereas the width is of the order of $10^{-8}$. The smallness rather holds to the way the resonance depends on $\alpha$. For instance, for pair (4,0), when $\alpha$ increases starting from $0$, the resonance follows the curve shown in Fig. 6 in the complex plane of variable $\zeta$.

\begin{figure}[h]
\centerline{\psfig{file=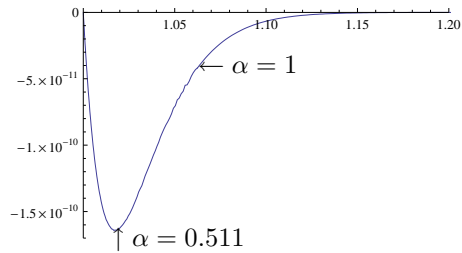,width=6cm}}
\vspace*{8pt}
\caption{Standard resonance for pair $(4,0)$. Variation of $\zeta$ with $\alpha$.\label{f6}} 
\end{figure}

\noindent
We thus see that increasing the coupling strength does not increase the resonance's width, but on the contrary diminishes it. We already met this phenomenon in such a two-level problem, when the coupling function is such that the standard resonance becomes a negative eigenvalue as the coupling increases (Ref. 10). Of course, in some situations, the resonance's width may also increase with the coupling (Fig. 8 of Ref. 10). It all depends on the coupling function of the problem that is considered (its shape, its width and the coupling strength).

\subsection{Conclusion}

In summary, we showed various types of quite narrow resonances (widths between $0$ and $50$ MeV), apart from the standard resonance which is always narrow in the case we considered: resonances $z_0$ for pairs $(n,n-1)$ (zero width) and pairs $(2,0),\ (3,1),\ (4,1)$ one the one hand; resonances $z_1$ for pairs $(3,2),\ (4,3),\ (2,0),\ (3,1)$ on the other hand. For other pairs for which the widths are larger, these widths may still be small in comparison with the energy difference between the two levels.

When the space extent of the pair increases, we see for each of the families: $(n,n-1)$, $(2n,0)$, $(2n+1,0)$, $(2n,1)$, $(2n+1,1)$, that resonance $z_1$ comes nearer to the lower level of the pair when $n$ increases. For all the series but the first one, if $n$ increases, the transition to the upper level is a transition from a fixed state to a state where the particle sits farther from its attraction center. For resonance $z_0$, except for the series $(n,n-1)$, the lowering also occurs but is only relative.

\section{General conclusion}

For the two systems we have been considering, the interest of taking account of \ns has been confirmed, as regards both characteristics of these resonances. Firstly their widths. We showed cases where they are small or vanish. Secondly their energies, which have a greater physical meaning when the width is small, absolutely or relatively to the gap between the two levels. We found energies comparable to those of the considered levels. For example, they may be below the lower level or between the two levels. When the resonances are narrow (respectively of zero width), one might say that the building of such a resonance (resp. a mixed mater-field state) by the field is in competition with the excitation of the particle up to the upper level.

We get a partial answer to the question raised in the Introduction about states appearing when the particles bound by certain potentials move apart from each other. We have seen how the excitation of the levels influence the energies of the \ns together with their widths. Such a relation had already been noticed in the case of the square-well potential studied in Ref. 3. Here, in several cases we noted that the larger the space extent of the pair, the smaller the gap between the energy of one of the \ns and the lower level of the pair. This space extension may be realized through the excitation of the only upper level or through the excitation of both levels. The former case corresponds to the particle moving away from the origin. This moving away is thus accompanied by the presence of a more and more bound (unstable) "state". In the harmonic oscillator case, the smallness of the resonances' widths  makes the phenomenon particularly interesting. For the $1/r$ potential, the connection between the extent of the pair and the lowering of the resonances' energy that we saw for the first values of $n$ is less interesting because of the resonances' width.

The two examples differ in that, when the excitation degree increases, the spacing of two consecutive levels diminishes in one case and is constant in the other case.
Let us point out an other difference between the two cases, as regards the nonstandard resonances' origin seen from a mathematical point of view. For the $1/r$ potential, the coupling functions are bounded at infinity but they have a pole in the lower half-plane. It is toward this pole that the \ns tend when $\alpha$ goes to $0$. For the harmonic oscillator, the coupling functions do not have any finite distance pole but are not bounded at infinity. The \ns then tend to infinity when the coupling to the field tends to $0$. The fact that there is several \ns for each pair is due in both cases to the fact that the singularity of the coupling function is not simple.

This study completes our preceding works on the interaction of a system with a massless boson field. We now have calculations of \ns in several cases: a particle in a square-well coupled to the boson (Ref. 3); a two-level atom coupled to the electromagnetic field in a cavity with various quality factors, the coupling being modeled by a Lorentzian function (Ref. 4); a particle in a $1/r$ potential weakly or strongly coupled to the boson (Ref. 4 and the present study); a harmonic oscillator weakly or strongly coupled to the boson (Ref. 5 and the present study); a particle in a  $ar+b/r$ potential strongly coupled to the boson (Ref. 6). We note that singularities of the coupling function at infinity play a role in both the square-well and the harmonic oscillator cases; therefore they are not to be neglected. We have not studied them in the $ar+b/r$-case.

In all the cases we studied, the calculations of the resonances required that we considered only two levels. In order to go beyond the qualitative results that this assumption has provided and to get quantitative predictions for a realistic quark-antiquark potential, it would be interesting to know how to tackle a three-level problem. This problem seems to be complicated, if only because of the involved multivalued analytic structure of the functions coming into play. We would like to know what becomes of the resonances we have constructed through artificially selecting two levels; the change in the structure of the set of resonances is likely to be important. Our simplification has provided some ideas to be tested in more elaborate models.

In the Introduction, we mentioned possible calculations with the $ar+b/r$ potential which contains a confining term. Another line of research might be to consider a potential without any confining force, such as the Yukawa potential, and determine whether there exist resonances of lower and lower energies analogous to the ones we saw in this paper and whether this could be related to a confinement mechanism.

\end{document}